# Low-Temperature Kinetics for the N + NO reaction: Experiment Guides the Way

Kevin M. Hickson,*,1 Juan Carlos San Vicente Veliz,2 Debasish Koner2,3 and Markus Meuwly*,2,4

1Univ. Bordeaux, CNRS, Bordeaux INP, ISM, UMR 5255, F-33400 Talence, France

2Department of Chemistry, University of Basel, Klingelbergstrasse 80, CH-4056 Basel, Switzerland.

3Department of Chemistry, Indian Institute of Technology Hyderabad, Sangareddy, Telangana 502285, India.

4Department of Chemistry, Brown University, Providence, RI 02912, USA

**Abstract**

The reaction N($^4$S) + NO(X$^2$Π) → O($^3$P) + N$_2$($X^1\Sigma_g^+$) plays a pivotal role in the conversion of atomic to molecular nitrogen in dense interstellar clouds and in the atmosphere. Here we report a joint experimental and computational investigation of the N + NO reaction with the aim of providing improved constraints on its low temperature reactivity. Thermal rates were measured over the 50 to 296 K range in a continuous supersonic flow reactor coupled with pulsed laser photolysis and laser induced fluorescence for the production and detection of N($^4$S) atoms, respectively. With decreasing temperature, the experimentally measured reaction rate was found to monotonously increase up to a value of (6.6 ± 1.3) × 10$^{-11}$ cm$^3$ s$^{-1}$ at 50 K. To confirm this finding, quasi-classical trajectory simulations were carried out on a previously validated, full-dimensional potential energy surface (PES). However, around 50 K the computed rates decreased which required re-evaluation of the reactive PES in the long-range part due to a small spurious barrier with height ~40 K in the entrance channel. By exploring different correction schemes the measured thermal rates can be adequately reproduced, displaying a clear negative temperature dependence over the entire temperature range. The possible astrochemical implications of an increased reaction rate at low temperature are also discussed.

# 1 Introduction

The reaction between atomic nitrogen in its ground electronic $^4S$ state, N($^4S$), and nitric oxide (NO)

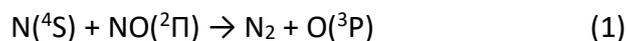
$$N(^4S) + NO(^2\Pi) \rightarrow N_2 + O(^3P) \qquad (1)$$

is a well-known process occurring in several planetary atmospheres such as the Earth,[1] Mars[2] and Venus.[3] The N + NO reaction is also part of the Zel'dovich reaction scheme relevant in combustion processes.[4] In the Earth's thermosphere, the reactions of NO with atomic nitrogen in its ground N($^4S$) and excited N($^2D$) state limit the total concentration of molecules with an odd number of nitrogen atoms by reforming molecular nitrogen, $N_2$. In the Martian upper atmosphere, reaction (1) is also the major loss process for odd nitrogen[2,5] while it also limits the extent of the night airglow due to NO in the Venusian atmosphere.[3] In dense interstellar clouds, reaction (1) plays an important role in the conversion of atomic to molecular nitrogen, with $N_2$ considered to be the major reservoir species for elemental nitrogen in the interstellar medium.[6] As the N≡N bond of $N_2$ is among the strongest of all covalent bonds, elemental nitrogen in this form is particularly difficult to liberate. Consequently, from the astrochemical modeling perspective, it is crucially important to have a good handle on the rate of reaction (1) at low temperature as this process constrains the amount of labile nitrogen available for further reaction and the eventual incorporation in complex organic molecules. However, despite its importance in interstellar chemistry, there are significant discrepancies between the results of previous experimental and theoretical studies at low temperature, making it difficult to establish the rates to be used for this process in computational modeling of chemical evolution interstellar models.

On the theoretical side, there are several studies of the kinetics of the N + NO reaction at low temperature employing different methodologies. These include quasi-classical trajectory calculations,[7-9] variational transition state theory[10] and quantum mechanical calculations.[11-13] According to these studies and at room temperature or below, reaction (1) occurs almost exclusively on the $^3A''$ potential energy surface (PES) of $N_2O$, with little or no contribution from the excited state $^3A'$ PES. The general consensus of these studies is that the reaction rate increases as the temperature falls between 300 K and 100 K followed by a decrease of the rate coefficient below 100 K.[8,11,12]

Although there are numerous previous experimental studies of the kinetics of the N + NO reaction at room temperature,[14-22] there are relatively few measurements at lower temperatures. Lee et al.[23] used two different experimental techniques: a discharge flow tube method (DF) and a flash photolysis system (FP) coupled with resonance fluorescence (RF) detection to measure the rate coefficients for reaction (1) over the 196-400 K range. Wennberg et al.[24] applied a similar DF-RF technique to the one used by Lee et al.[23] to study this reaction over the 213-369 K temperature range. More recently, Bergeat et al.[25] measured the rates of the N + NO reaction between 48 and 211 K using a DF method for the production of atomic nitrogen coupled with a continuous supersonic flow reactor to reach low temperatures. A similar RF detection method to the ones applied by Wennberg et al.[24] and Lee et al.[23] was used to follow the kinetics of atomic nitrogen disappearance. Nevertheless, as a large fraction of the atomic nitrogen was consumed upstream of the cold supersonic flow in the work of Bergeat et al.[25] (60-98 % of the N($^4$S) produced upstream of the reactor by the microwave discharge of $N_2$ was lost before reaching the cold flow), the potential for error was significant in these experiments. Interestingly, the measurements of Bergeat et al.[25] showed that the rate coefficients for this reaction continue to increase down to 48 K, reaching a value of $(5.8 \pm 0.3) \times 10^{-11}$ cm$^3$ s$^{-1}$ at 48 K, in contradiction with the results of the majority of theoretical studies.[8, 11, 12] Indeed, if the Bergeat et al.[25] data is extrapolated to 10 K using an Arrhenius type extrapolation, a rate coefficient greater than $3 \times 10^{-10}$ cm$^3$ s$^{-1}$ is obtained, in stark contrast to the (calculated) values which are currently used as the basis for the rate coefficients of this reaction in astrochemical models (the recommended rate coefficient for reaction (1) in the KIDA database is $1 \times 10^{-11}$ cm$^3$ s$^{-1}$ cm$^3$ s$^{-1}$ at 10 K).[26]

This large discrepancy could have important consequences for astrochemical models of nitrogen-bearing organic species in interstellar clouds. If the rate coefficient for reaction (1) is large, as determined by Bergeat et al.,[25] elemental nitrogen would be mostly locked up in the form of $N_2$ in interstellar clouds and therefore unavailable for further reaction. Conversely, if the majority of theoretical predictions are correct, significantly more nitrogen would be available in a labile form.

Given the differences between theory and the results of a single experimental measurement close to astrochemically relevant temperatures, there is evidently the need for further experimental and/or theoretical confirmation to elucidate the behaviour of the N + NO reaction as the temperature falls. In order to address this issue, we report the results of a new experimental investigation of reaction (1) down to 50 K using the same continuous supersonic flow apparatus as used by Bergeat et al.[25] In this instance, however, entirely different techniques have been employed for the in-situ production and pulsed detection of ground state atomic nitrogen atoms. In previous work,[25] it was not possible to generate nitrogen atoms in-situ due to the use of microwave discharge methods for their production and detection. To support the experimental findings at low temperatures, quasi-classical trajectory (QCT) simulations were carried out on a validated, full-dimensional PES. It was found that the results are very sensitive to details of the PES in the region where the impacting nitrogen atom can approach the collision partner at low temperatures and additional refinement of the PES yields improved results.

The present work is structured as follows. First, the experimental and theoretical methods used are described in sections 2 and 3, respectively. The results are presented in section 4 and the discussion and conclusions of this work are given in section 5.

## 2 Experimental methods

The present measurements were performed on an existing supersonic flow reactor, also known as the CRESU technique (CRESU is a French acronym for Cinétique de Réaction en Écoulement Supersonique Uniforme or reaction kinetics in a uniform supersonic flow). A detailed description can be found in some of the earliest papers employing this apparatus,[27,28] and will not be repeated here. This particular CRESU reactor operates in continuous mode, employing axisymmetric Laval nozzles that generate a cold supersonic flow through the isentropic expansion of a carrier gas from a higher-pressure region into the evacuated reactor downstream of the nozzle. The convergent-divergent nozzle profile is specifically designed to produce a downstream flow with uniform density, temperature and velocity as a function of distance from the nozzle. In this way, the cold flow generated by the nozzle represents a suitable environment to study the kinetics of fast gas-phase reactions at low temperature. The various modifications that have been applied to facilitate this work will be described below alongside details of the methodology specific to this investigation. During the present study, three different Laval nozzles based on either Ar or $N_2$ as carrier gases were used to generate four different low temperature flows (one nozzle was used with both Ar and $N_2$) of 177, 127, 75 and 50 K. Additional measurements were performed at room temperature, 296 K, without a Laval nozzle effectively using the reactor as a slow-flow system.

In our previous study of the $N(^4S)$ + NO reaction,[25] $N(^4S)$ atoms were generated by the microwave discharge technique, using a cavity operating at 2.45 GHz to dissociate molecular nitrogen upstream of the Laval nozzle. While this method allows large N($^4$S) concentrations to be generated continuously, the use of this source also presented a couple of major disadvantages. First, as mixing of the reagent gases upstream of the reactor was unavoidable due to the nature of the Laval nozzle method (any attempt to inject reagents downstream of the nozzle results in the destruction of the supersonic flow) an overwhelming fraction of the initial N($^4$S) atoms was removed by reaction with NO before reaching the cold flow. The problem was exacerbated by the long residence times of the gas in the upstream region and Laval nozzle reservoir. Second, the use of a continuous source of atomic radicals required us to physically move the Laval nozzle with respect to the observation axis to modify the reaction time. This procedure leads to an additional source of error in the experiments due to the slight variations in density, temperature and velocity of the gas flow as a function of distance which are always present in supersonic flows generated by Laval nozzle expansions.

Ideally, a pulsed photolytic source of N($^4$S) atoms would be used which would allow these atoms to be produced directly in the cold flow. Furthermore, the use of a pulsed source (coupled with a pulsed detection method) avoids the requirement to displace the nozzle as the timing is controlled by the delay between the two lasers. In the present work, ground state atomic nitrogen was produced by chemical reaction following the pulsed photolysis of the precursor molecule CBr$_4$, in a similar manner to the study of the N($^2$D) + NO reaction by Nuñez-Reyes et al. [29]

Here, C($^3$P) atoms were initially produced by CBr$_4$ photolysis at 266 nm using a pulsed (10 Hz) frequency quadrupled Nd:YAG laser with approximately 30 mJ pulse energy and a 5 mm diameter beam that was aligned along the length of the supersonic flow. CBr$_4$ itself was introduced into the flow by passing a small fraction of the carrier gas into a flask containing solid CBr$_4$ maintained at a fixed known temperature and pressure. C($^3$P) atoms were allowed to react with NO, a process with several different barrierless exothermic product channels

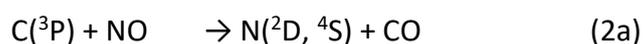
$$C(^3P) + NO \rightarrow N(^2D, ^4S) + CO \qquad (2a)$$
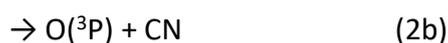
$$\rightarrow O(^3P) + CN \qquad (2b)$$

The room temperature branching ratio for the two product channels (2a)/(2b) has been estimated to be 1.5 ± 0.3[30] with a total rate coefficient (channels 2a+2b) of 1.5 × 10$^{-10}$ cm$^3$ s$^{-1}$ at 298 K increasing to 2.0 × 10$^{-10}$ cm$^3$ s$^{-1}$ at 50 K although the relative proportions of N($^4$S) and N($^2$D) formed by this reaction are unknown [31] but would be amenable to computation.[32]

The previous theoretical work of Andersson et al. [33] indicates that the channel forming atomic nitrogen should decrease as the temperature falls.

N($^4$S) atoms were detected directly during the present work by pulsed laser induced fluorescence in the vacuum ultraviolet range through the $2s^22p^3$ $^4S°_{3/2}$ - $2s^22p^2(^3P)3s$ $^4P_{1/2}$ transition at 120.071 nm. For this purpose, the tunable visible output of Nd:YAG pumped dye laser around 720.4 nm was frequency doubled in a BBO crystal, producing UV radiation around 360.2 nm with a pulse energy around 9 mJ. After separation of the residual visible light through two dichroic mirrors optimized for reflection around 355 nm, the UV beam was steered and focused into a cell containing 800 Torr of Kr to generate the VUV beam by frequency tripling. A MgF$_2$ lens acted as the cell exit window to collimate the VUV beam while simultaneously leaving the residual UV beam to diverge due to the different refractive indices of MgF$_2$ at UV and VUV wavelengths. The cell itself was positioned at the end of a long sidearm (75 cm) containing circular diaphragms, to further reduce the intensity of the UV beam reaching the reactor. The sidearm was attached to the reactor at the level of the observation region. The zone between the cell exit window and the reactor was continuously flushed by Ar or N$_2$ to prevent supplementary VUV absorption losses. Resonant emission from N($^4$S) atoms within the cold flow was detected at right angles to both the excitation and flow axes. Here, a LiF lens was used to focus the VUV emission onto the photocathode of a solar blind photomultiplier tube (PMT). A LiF window was used to isolate the reactor from the lens and the PMT to protect these elements from reactive gases, while the region between the PMT and the LiF window was maintained at low pressure (10$^{-2}$ mbar) to prevent atmospheric absorption losses. Although VUV radiation was unambiguously generated, as shown by the detection of fluorescence from N($^4$S) atoms only when Kr was present in the tripling cell, the generation efficiency was very low compared with previous experiments at other VUV wavelengths. [29, 34-36] This was not unexpected given the previous work by Mahon et al. [37] clearly showing that the tripling efficiency becomes negligibly small at this wavelength. Indeed, when the probe laser was tuned to the N($^4$S) transition at the slightly shorter wavelength of 120.0223 nm, little or no fluorescence signal could be observed. The PMT output was fed into a fast preamplifier before passing into a gated integrator. The delay between photolysis and probe laser pulses was controlled by a digital pulse generator which also triggered the acquisition electronics and oscilloscope for signal monitoring purposes. A minimum of 100 time points was recorded to establish the decay profiles, with each time point being generated by the average of 30 individual laser shots. Baseline levels were set by recording the gated signal levels at negative time delays.

The gases used in these experiments were regulated by calibrated mass-flow controllers and taken directly from cylinders with the following purities (N$_2$ 99.995 %, Ar 99.999 %, NO 99.5 %, Kr 99.999 %).

**3 Computational Methods**

There are two electronic states ($^3$A' and $^3$A'') that connect the N($^4$S) + NO(X$^2$Π) and O($^3$P) + N$_2$($X^1\Sigma_g^+$) asymptotes. At low temperatures (T ≤ 300 K) the contribution to the total reaction rate from the $^3$A' electronic state is vanishingly small due to an inner barrier (~ 4820 K) in the entrance channel, whereas the $^3$A'' PES exhibits a barrierless path for this reaction and contributes entirely to the reaction at low temperatures. This was already found in previous calculations[9] using a reproducing kernel Hilbert space (RKHS) representation of reference *ab initio* energies at the MRCI+Q/aug-cc-pVTZ level of theory. This PES is referred to as PES2020 in the following.

QCT simulations using PES2020 for the $^3$A'' PES lead to decreasing rates below approximately 75 K. At low temperatures, access of the impacting nitrogen atom N($^4$S) for reactive collisions with NO(X$^2$Π) is along Jacobi angles $\theta \in [110 - 135]^o$ where θ is the angle between the NO bond vector and the vector *R* (distance between the center of mass of NO and the nitrogen atom). Careful analysis of the PES revealed a small barrier (~40 K) in the RKHS based on the MRCI+Q reference calculations along the cut θ=129.967° with the maximum at *R* = 12.90 a$_0$. This spurious barrier arises due to scarcity of the reference data in the long-range part of the PES which originates also from difficulties in converging the reference MRCI+Q calculations with large basis sets in these regions of the PES. This was ascertained from CASSCF and CASPT2 calculations with the same basis set which did not feature such a barrier.

Because the origin of the limitation is known, two different means to account for and to correct the erroneous behaviour were explored. They involve replacing the MRCI+Q reference energies in the long range (*R* > 7a$_0$) by either explicit long-range expansions which is an extensively used approach,[38-41] or by extrapolating the converged MRCI+Q energies using an energy expression with the leading long-range term proportional to 1/R$^6$. This is followed by reconstructing the RKHS representation after completing the grid for *r* = 2.185 a$_0$ which is the equilibrium bond length of NO. The same procedure was applied to all cuts of $r$ and the 3-dimensional RKHS was reconstructed.

Because N($^4$S) is essentially spherically symmetric it has no low-order permanent electrical moments and the leading contribution in the long range is the attractive part of the van der Waals interaction $V(R) = 4\varepsilon_{ij}\left[\left(\frac{\sigma_{ij}}{R}\right)^{12} - \left(\frac{\sigma_{ij}}{R}\right)^{6}\right]$ between atoms $i$ and $j$. In the following, atomic parameters ($\varepsilon_N$ = 0.20 kcal/mol; $\sigma_N$ = 2.00 Å) and ($\varepsilon_O$ = 0.16 kcal/mol; $\sigma_O$ = 2.05 Å) for nitrogen and oxygen were employed.[42] Using Lorentz-Berthelot combination rules ($\sigma_{ij} = \sqrt{\sigma_i \sigma_j}$ and $\varepsilon_{ij} = \frac{1}{2}(\varepsilon_i + \varepsilon_j)$) the ratios $\frac{V_{LJ}(R=9)}{V_{LJ}(R=7)}$ and $\frac{V_{LJ}(R=12)}{V_{LJ}(R=7)}$ are 0.2168 and 0.0380, respectively, from which scaling factors for obtaining energies at $R$ = 9 and $R$ = 12 $a_0$, respectively, were determined from the converged MRCI+Q energies at 7 $a_0$. The distances $R$ = 9 and $R$ = 12 $a_0$ are chosen to cover regions over which the long-range interaction decays significantly compared with $V(R = 7)$. This allows to complete the grid and reconstruct the full dimensional PES using RKHS theory which is referred to as PES2022-LJ1. In order to assess the sensitivity of the results on this approach, a second PES (PES2022-LJ2) was determined by reducing the van der Waals interaction strengths $\varepsilon_{ij}$ by a factor of two.

As a second possibility, reference energies beyond $R$ = 7 $a_0$ were determined based on the leading long-range term ~$1/R^6$ for neutral—neutral (N—NO) interactions by extrapolation. Since a $k^{[2,6]}$ reciprocal kernel decays smoothly following $1/R^6$ as the leading term,[43] the converged MRCI+Q energies (up to and including $R$ = 7 $a_0$) for $r$ = 2.185 $a_0$ only and for all values of $\theta$ were extrapolated on a grid along $R$ within the RKHS formalism using $k^{[2,6]}(R)$. The extrapolated energies were then used for each value of $\theta$ to replace the unconverged MRCI+Q energies along $R$ for $r$ = 2.185 $a_0$. A subsequent, 2D RKHS interpolation was performed to obtain $V(R, r)$ for each Jacobi angle, to complete the energy grid for a specific angle. $V(R, r)$ was expressed as the product of two reciprocal 1D kernels ($k^{[2,6]}(R)$). Since the 1D reciprocal kernel approaches zero at large distance, it is necessary to subtract $V(R \to \infty)$. A 3D surface was then obtained by using the completed energy grid for all angles via 3D RKHS interpolation for each channel (N + NO, O + N$_2$). Finally, the global reactive 3D surface, PES2022-R, was constructed by mixing the PESs of the two channels, see Ref. 9.

**4 Results**

## 4.1 Experimental results

To determine the experimental rate coefficients, the pseudo-first-order approximation was applied during the present investigation, through the use of low concentrations of C($^3$P) atoms with respect to the large excess concentration of NO. Under these conditions, N($^4$S) atoms are initially formed by reaction (2a), before being consumed by reaction with NO, following a biexponential rate law of the type

$$[N(^4S)] = A(exp(-k'_a t) - exp(-k'_b t)) \qquad (3)$$

where $[N(^4S)]$ is the time dependent concentration of N($^4$S) atoms, which is considered to be proportional to the N($^4$S) fluorescence intensity, A is a coefficient, $k'_a$ is the pseudo-first-order rate coefficient for N($^4$S) loss, $k'_b$ is the pseudo-first-order rate coefficient for N($^4$S) formation and $t$ is the time. As the VUV probe radiation was much weaker than used in previous studies, it was necessary to use high CBr$_4$ concentrations and high photolysis laser pulse energies, while a preamplifier was used on the PMT output to increase the N($^4$S) signal levels. An unfortunate side effect of these modifications was an increased scattering of the UV photolysis beam off the large CBr$_4$ molecules and onto the PMT. Despite the use of a solar bind type PMT, this scattering effect resulted in saturation of the PMT for 15 microseconds following the photolysis laser pulse. Consequently, as it was impossible to record a large part of the initial rise of the fluorescence profile, a single exponential decay function of the form

$$[N(^4S)] = A exp(-k'_a t) \qquad (4)$$

was used instead to fit to the fluorescence signals as shown in Figure 1 for experiments performed at 50 and 296 K.

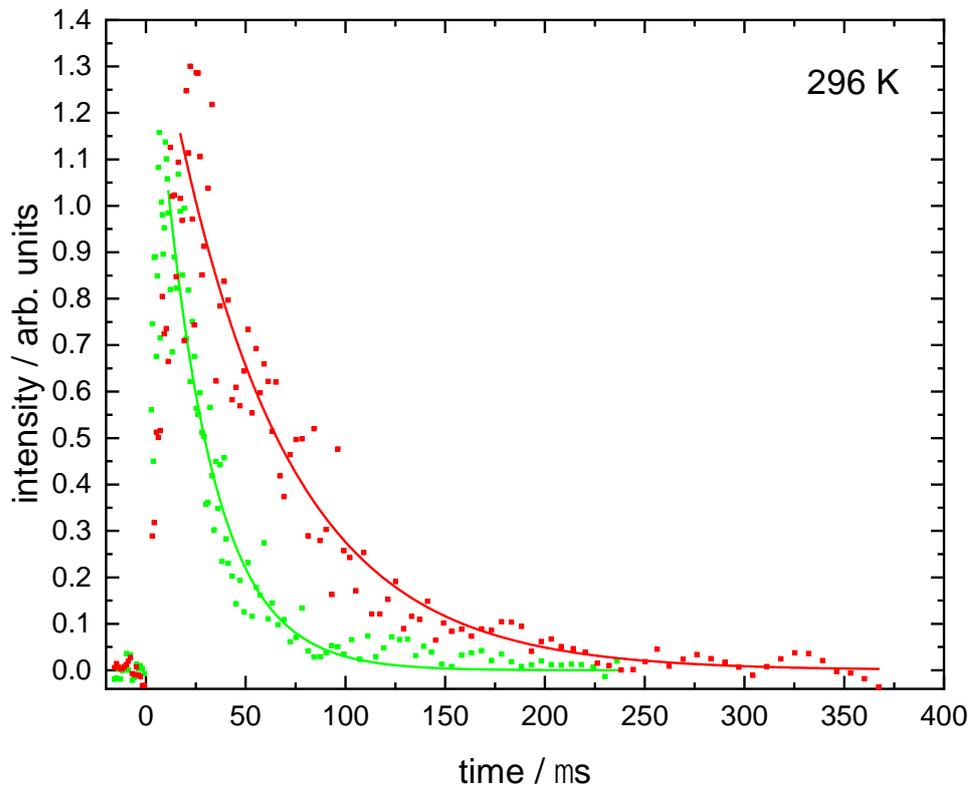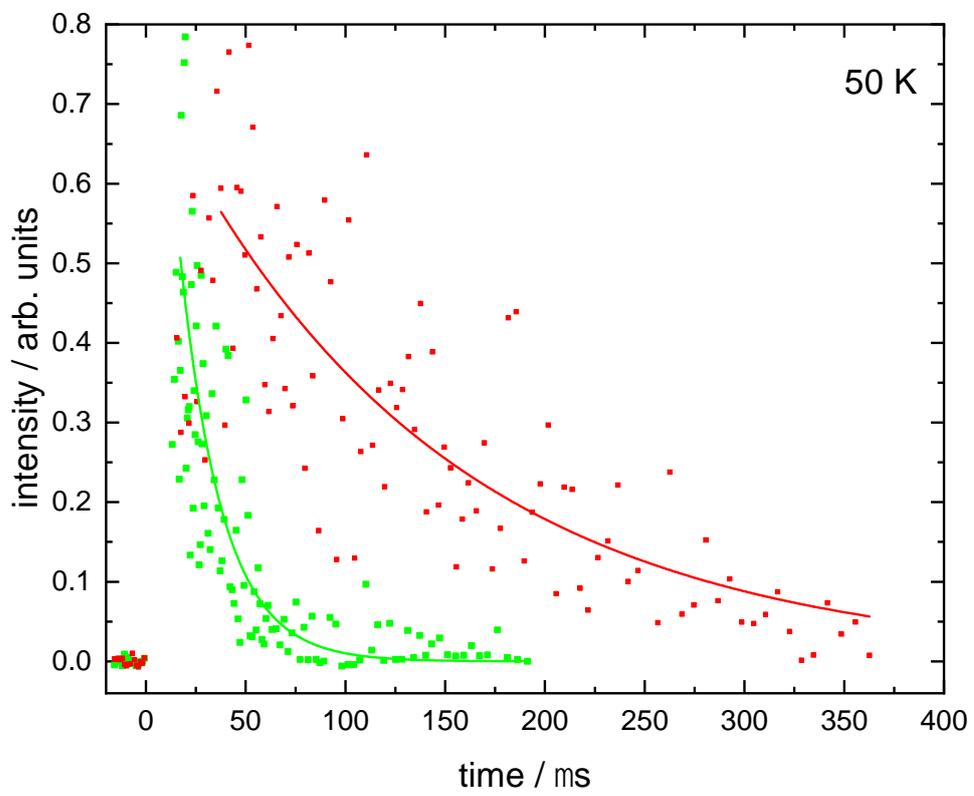

**Figure 1** N($^4$S) fluorescence emission as a function of time in the presence of NO. **Upper panel**: Data recorded at 296 K. (Red solid circles) [NO] = (6.3 ± 0.1) × 10$^{14}$ cm$^{-3}$; (green solid circles) [NO] = (14.6 ± 0.1) × 10$^{14}$ cm$^{-3}$. **Lower panel**: Data recorded at 50 K. (Red solid circles) [NO] = (10.8 ± 0.4) × 10$^{13}$ cm$^{-3}$; (green solid circles) [NO] = (63.5 ± 2.4) × 10$^{13}$ cm$^{-3}$. Solid red and green lines represent single exponential fits to the data.

Here, the starting point of the analysis had to be carefully chosen to avoid fitting to the rising part of the N($^4$S) fluorescence signal as described by the second term of expression (3). This starting point was generally delayed by at least several microseconds with respect to $t = 0$ as shown in Figure 1 and was selected by plotting the fluorescence signals on a logarithmic scale. Then only those datapoints that obeyed the simple exponential loss expression (4) (i.e. displaying a 'linear' dependence on the logarithmic scale) were used to derive $k'_a$. Fluorescence decays were recorded for a minimum of seven different NO concentrations at each temperature. By plotting the derived $k'_a$ values as a function of [NO], the second-order rate coefficients could be determined from the slope of weighted linear-least squares fits to the data. These fits and the associated data obtained at all temperatures (50, 75, 127, 177 and 296 K) are shown in Figure 2.

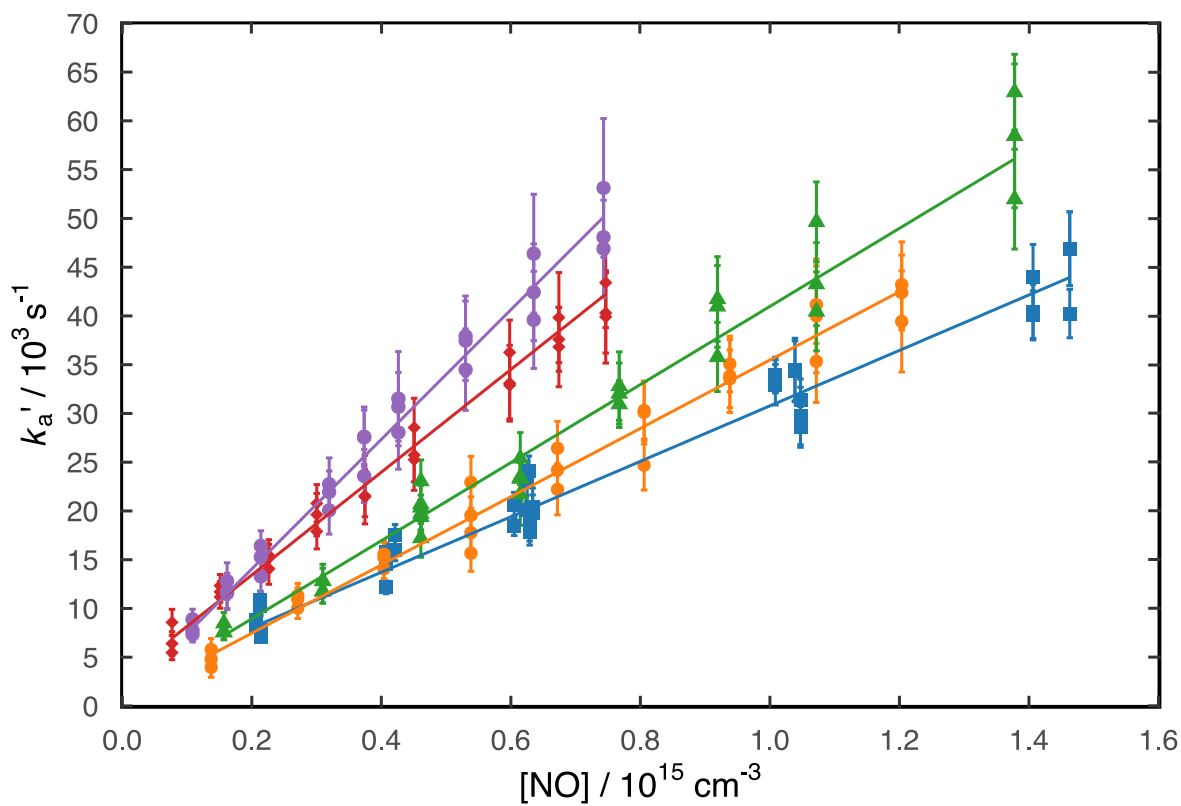

**Figure 2** Measured pseudo-first-order rate coefficients as a function of NO concentration recorded at different temperatures. (Blue squares) 296 K; (orange circles) 177 K; (green triangles) 127 K; (red diamonds) 75 K; (purple circles) 50 K. Solid lines represent weighted fits to the individual datasets with the slopes of these fits yielding the second-order rate coefficients.

These data are summarized in Table 1 alongside other relevant information and are shown as a function of temperature in Figure 3 together with earlier experimental and theoretical work.

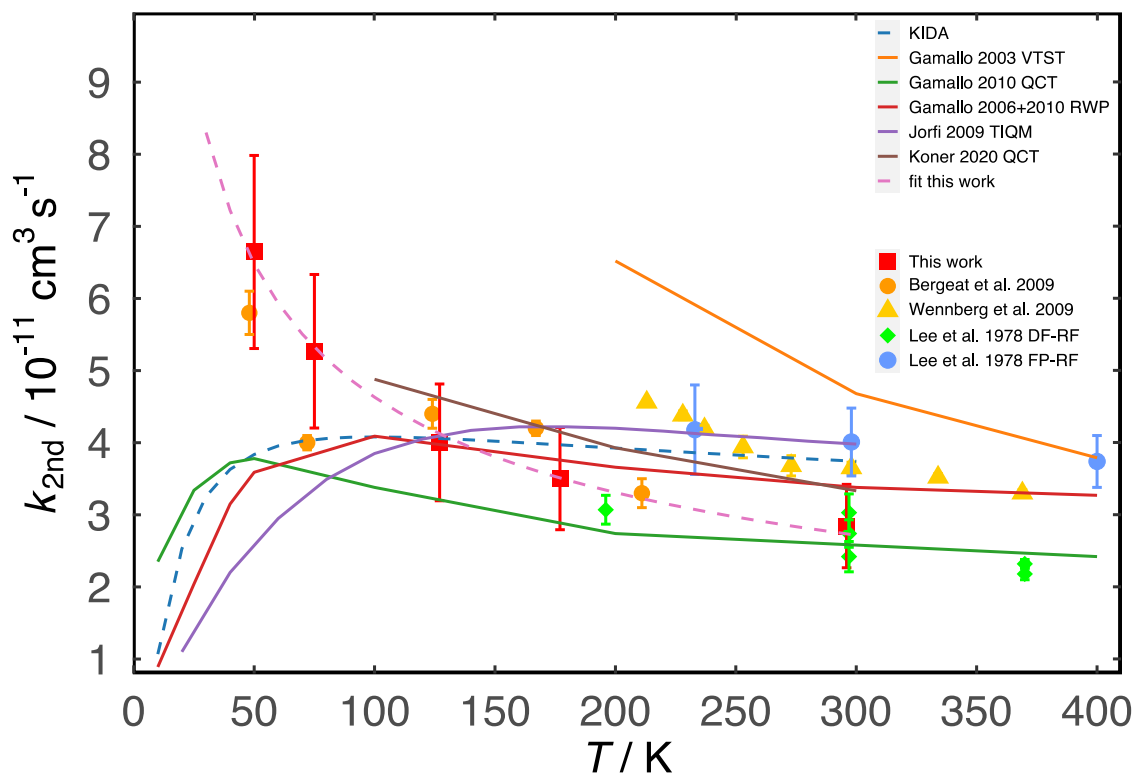

**Figure 3** Temperature dependent rate coefficients for the N($^4$S) + NO reaction. **Experiments** (green diamonds) Lee et al.,[23] discharge flow – resonance fluorescence method; (blue circles) Lee et al.,[23] flash photolysis – resonance fluorescence method; (yellow triangles) Wennberg et al.[24]; (orange circles) Bergeat et al.[25]; (red squares) this work; (dashed purple line) fit to this work using the expression $k = (2.7 \pm 0.2) \times 10^{-11} (T/300)^{-(0.50 \pm 0.02)}$ cm$^3$ s$^{-1}$. **Theory** (orange line) Gamallo et al.,[10] variational transition state theory; (red line) Gamallo et al.,[11] time dependent wavepacket calculations; (green line) Gamallo et al.,[8] quasi-classical trajectory calculations; (purple line) Jorfi and Honvault,[12] time-independent quantum mechanical method; (brown line) Koner et al.,[9] quasi-classical trajectory calculations. (Dashed blue line) currently recommended rate coefficients from KIDA (KInetic Database for Astrochemistry).[26] Lines represent fits to parametrized expressions.

**Table 1** Temperature dependent rate coefficients for the N($^4$S) + NO reaction

| T / K | $N^b$ | Flow density / 10$^{16}$ cm$^{-3}$ | [NO]/ 10$^{14}$ cm$^{-3}$ | $k_{N(^4S)+NO}$ / cm$^3$ s$^{-1}$ |
| --- | --- | --- | --- | --- |
| 296 | 23 | 16.3 ± 0.2 | 2.1-14.6 | (2.8 ± 0.6)[c] × 10$^{-11}$ |
| 177 ± 2 [a] | 28 | 9.4 ± 0.2 | 1.4-12.0 | (3.5 ± 0.7) × 10$^{-11}$ |
| 127 ± 2 | 29 | 12.6 ± 0.3 | 1.6-13.8 | (4.0 ± 0.8) × 10$^{-11}$ |
| 75 ± 2 | 27 | 14.7 ± 0.6 | 0.8-7.5 | (5.3 ± 1.1) × 10$^{-11}$ |
| 50 ± 1 | 27 | 25.9 ± 0.9 | 1.1-7.4 | (6.6 ± 1.3) × 10$^{-11}$ |

[a]Errors associated with the calculated temperatures represent the statistical (1σ) uncertainites obtained from Pitot tube measurements of the impact pressure. [b]Number of individual measurements. [c]Uncertainties on the measured rate coefficients represent the statistical errors (1σ) derived from weighted fits such as those shown in Figure 2, combined with an estimated systematic error of 20%.

One possible source of error in the present experiments arises due to the presence of N($^2$D) atoms in the flow, produced by reaction (2a) alongside N($^4$S) atoms. During these experiments, N($^2$D) atoms will also be removed by NO with rate coefficients that are between 1.7 and 2.6 times faster than the rate coefficients for the ground state nitrogen atom reactions at the same temperatures. In this case, N($^2$D) removal could occur through either reactive collisions which would not interfere with our kinetic measurements of the N($^4$S) + NO reaction, or by quenching collisions leading to N($^4$S) atom formation, which could potentially alter the measured kinetic decays. N($^2$D) quenching by the carrier gases N$_2$ and Ar has already been shown to be negligibly small over the timescales of the present experiments. [44, 45] In order to examine the possible interferences due to N($^2$D) quenching by NO, an additional experiment was performed at 296 K, where a large excess H$_2$ concentration of 1.2 × 10$^{16}$ cm$^{-3}$ was added to the flow. The room temperature rate coefficient for the N($^2$D) + H$_2$ reaction was recently measured with a value of $k_{N(^2D)+H_2}$(296 K) = (2.2 ± 0.5) × 10$^{-12}$ cm$^3$ s$^{-1}$. [46] Although there are several other coreagents that could be employed to remove N($^2$D) atoms more rapidly than H$_2$, such as C$_2$H$_2$[47] and C$_2$H$_4$,[48] both of these species also react rapidly with the N($^4$S) atom precursor, C($^3$P) making it very difficult to detect N($^4$S) atoms in the presence of these reagents due to the low N($^4$S) concentrations generated. In contrast, C($^3$P) atoms do not react with H$_2$ in the gas-phase so that N($^4$S) signal levels are relatively unaffected. Under these conditions, the N($^2$D) atoms were removed by both the N($^2$D) + NO and N($^2$D) + H$_2$ reactions without leading to N($^4$S) atom production with calculated pseudo-first-order rate coefficients between 41000 s$^{-1}$ and 128000 s$^{-1}$ (this value varies due to the changing NO concentration), corresponding to half-life times for N($^2$D) loss of between 17 μs and 5 μs. The second-order rate coefficient derived during this test experiment of (2.9 ± 0.3) × 10$^{-11}$ cm$^3$ s$^{-1}$ is in excellent agreement with those experiments performed without H$_2$ of (2.8 ± 0.3) × 10$^{-11}$ cm$^3$ s$^{-1}$, demonstrating the negligible influence of N($^2$D) quenching on the kinetic results.

The room temperature rate coefficient obtained here ($k_{\mathrm{N(^4S)+NO}}$(296 K) = (2.8 ± 0.6) × 10$^{-11}$ cm$^3$ s$^{-1}$), based on the data obtained both with and without H$_2$, is seen to be in excellent agreement with the discharge-flow – resonance fluorescence results of Lee et al. [23] ($k_{\mathrm{N(^4S)+NO}}$(297 K) = (2.42-3.03) × 10$^{-11}$ cm$^3$ s$^{-1}$) and slightly lower than the values obtained by the flash photolysis - resonance fluorescence study of Lee et al., [23] ($k_{\mathrm{N(^4S)+NO}}$(298 K) = (4.01 ± 0.47) × 10$^{-11}$ cm$^3$ s$^{-1}$) and the discharge-flow tube study of Wennberg et al. [24] ($k_{\mathrm{N(^4S)+NO}}$(298 K) = (3.65 ± 0.08) × 10$^{-11}$ cm$^3$ s$^{-1}$). Despite this, there are numerous other room temperature studies of this reaction with rate coefficients in the range (1.7-4.5) × 10$^{-11}$ cm$^3$ s$^{-1}$ (see Table 1 of Nakayama et al. [22]) with an average value of 2.9 × 10$^{-11}$ cm$^3$ s$^{-1}$; in excellent agreement with the present result.

Below room temperature, the present results are in good agreement with the earlier study of Bergeat et al. [25] who employed the same continuous flow CRESU apparatus, but with entirely different N($^4$S) production and detection methods. While some of the rate coefficients derived by Bergeat et al. [25] do not display a uniform progression as a function of temperature, particularly over the range 75-177 K, the overall trend is still very similar to the one shown by these new measurements if the values obtained at higher (211 K and 296 K) and lower (50 K) temperature are also taken into consideration. The present results show a clear tendency towards an acceleration of the reaction rate as the temperature falls, reaching a maximum value of $k_{\mathrm{N(^4S)+NO}}$(50 K) = (6.6 ± 1.3) × 10$^{-11}$ cm$^3$ s$^{-1}$.

## 4.2 Computational Results

For the empirically corrected PESs - PES2022-LJ1, PES2022-LJ2, and PES2022-R – thermal rates at $T$ = 30, 50, 70, 100, 125, 175, 200, and 400 K were determined from QCT simulations, see

Table S1. For each temperature, a total of $N_{tot}$ = 500000 (for PES2022-LJ1 and PES2022-LJ2 to determine error bars) and 50000 (for PES2022-R) trajectories was run. The rate coefficient was determined from

$$k(T) = g(T)\sqrt{\frac{8k_bT}{\pi\mu}}\pi b_{max}^2 P \qquad (5)$$

where $P = \frac{N_{reac}}{N_{tot}}$ is the ratio between the number of trajectories that reacted ($N_{reac}$) and the total number of trajectories $N_{tot}$, $b_{max}$ is the maximum impact parameter, μ is the reduced mass and the degeneracy factor is

$$g(T) = \frac{3}{4 \times (2 + 2\,exp^{\frac{-177.1}{T}})} \qquad (6)$$

where the spin-orbit splitting for NO(X $^2\Pi_{1/2}$) and NO(X $^2\Pi_{3/2}$) is 177.1 K.[49] Stratified sampling[50] with 6 strata was used to sample the impact parameter *b* for different trajectories. It was found that ~40-45% trajectories were reactive at each temperature which guarantees convergence of the rate coefficients.

The results for the three new PESs are reported in Figure 4. The thermal rates for PES2020 show agreement with the current experimental results down to *T* = 75 K where it plateaus and decreases for lower temperatures. In contrast, results from PES2022-R and the two van der Waals adapted PESs (PES2022-LJ1 and PES2022-LJ2) keep on increasing for temperatures below 50 K and show qualitative agreement at all temperatures with the present measurements and those from the earlier work of Bergeat et al.[25] A similar increase in the rate towards lower temperatures (15 K) was found for the C($^3$P) + O$_2$($^3\Sigma_g^+$) forward reaction that connects to the O($^1$D) product electronic state. The product QCT rates[51] showed good agreement with previous experimental results.[52]

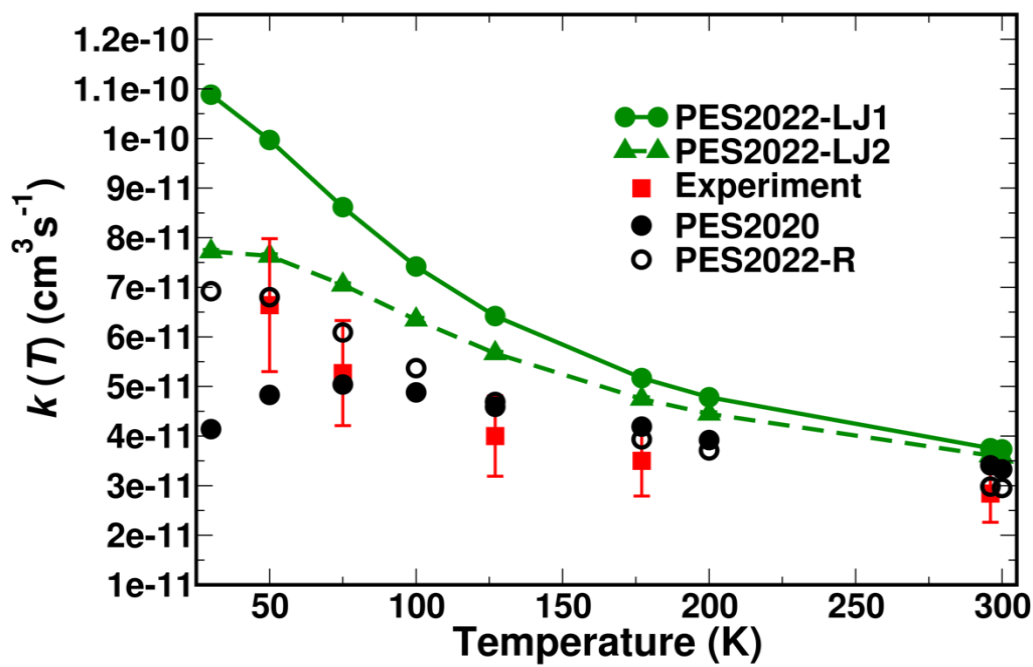

**Figure 4** Thermal rates for the forward reaction N($^4$S) + NO(X$^2$Π) → O($^3$P) + N$_2$($X^1\Sigma_g^+$) on the $^3$A'' PES. QCT simulations were carried out on PES2020 (black solid circle), PES2022-LJ1 (green solid circle), PES2022-LJ2 (green solid triangle), and PES2022-R (black open circle), together with the present results from experiments (red solid square together with error bars), see Table 1 for details of the error estimation. Analysis of the products from histogram and Gaussian binning[53] on PES2022-R found comparable results, see Table S1. For the simulations including PES-2022-LJ1/LJ2 500 000 trajectories were run from which error bars were determined by randomly sampling 50 000 trajectories via bootstrapping. The error for both PES-2022-LJ1/LJ2 was smaller than the symbol sizes.

To explore the link between the shape of the PES and the underlying reactive and non-reactive collisions, a number of trajectories are graphically represented in Figure 5.

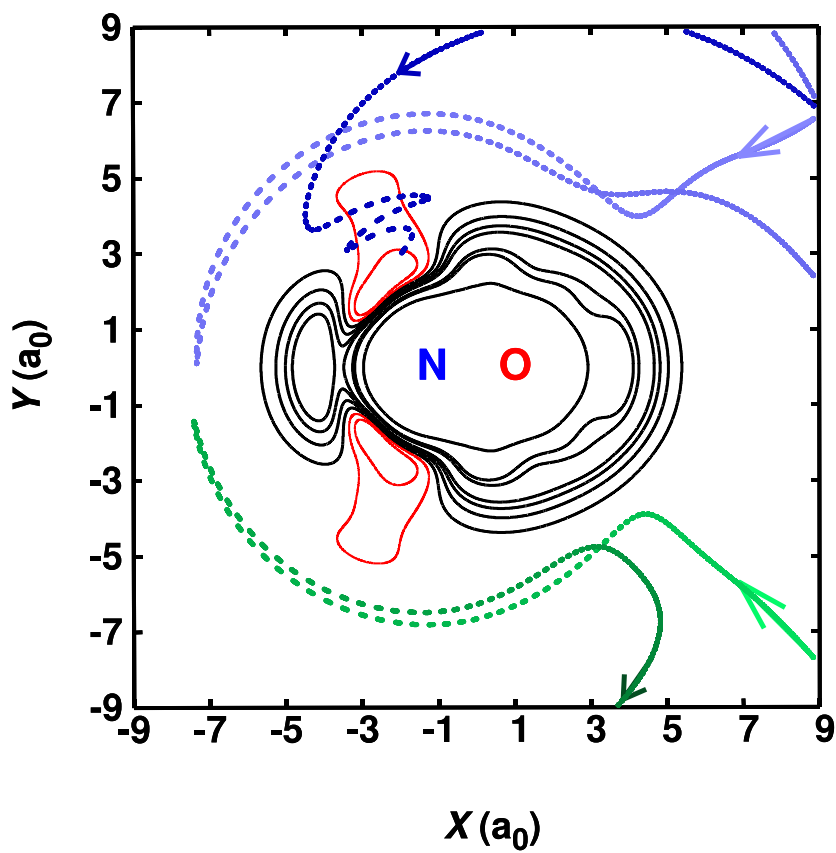

**Figure 5** Contour plot (solid lines) of the RKHS evaluated PES-2022-LJ1 energy surface at fixed equilibrium N-O separation of $r_{NO}$=2.185 $a_0$. Selected energy isocontours are shown for attractive [-0.3, -0.05] eV (red solid line) and repulsive [0.20, 0.45, 0.71, 0.96, 2.0, 2.5, 6.7] eV (black solid line) energies with respect to the N + NO ($r_{eq}$) energy. Additionally, results from QCT simulations are overlapped for one reactive (blue) and one non-reactive (green) trajectory. Reactive trajectories (light color for early and dark colors for late simulation times) are found to access the narrow funnel at θ = 129.0º leading to the minimum energy structure whereas nonreactive trajectories bypass this region.

The results displayed in Figure 5 are reported from simulations using PES2022-LJ2. It was found that reactive trajectories (blue) invariably sample the narrow access around θ = 129º for the nitrogen atom approaching the diatomic. For directions that deviate from this angle, the collision is elastic (green). This explains why at low temperatures any barrier at values of $R$ greater than ~ 8 $a_0$ leads to a pronounced decrease in the rate coefficient which is inconsistent with experiment, as was found for PES2020 (see also Figure S3). This, together with the results from CASSCF and CASPT2 calculations clarifies that the small barrier (see Figure S2) at long-range that is present in the RKHS is spurious and due to limitations in the quantum chemical method (MRCI+Q). It is also shown in Figure S2 that only the approach along θ ~129° suffers from this artifact.

Figure 6 compares the present PESs (PES2022-LJ1, PES2022-LJ2, PES2022-R) with the previous one (PES2020) for N approaching NO (with the NO bond distance at its equilibrium value of $r$ = 2.185 $a_0$) at an angle θ = 129.0°. The spurious barrier in PES2020 at 12.5 $a_0$ is evident whereas the other three PESs decay monotonically to zero for large values of $R$. The right-hand label of the $y$-axis provides energies in temperature units which clarifies that the height of the barrier (~40 K) corresponds to the simulation temperature at which the thermal rate decreases when going to lower temperature.

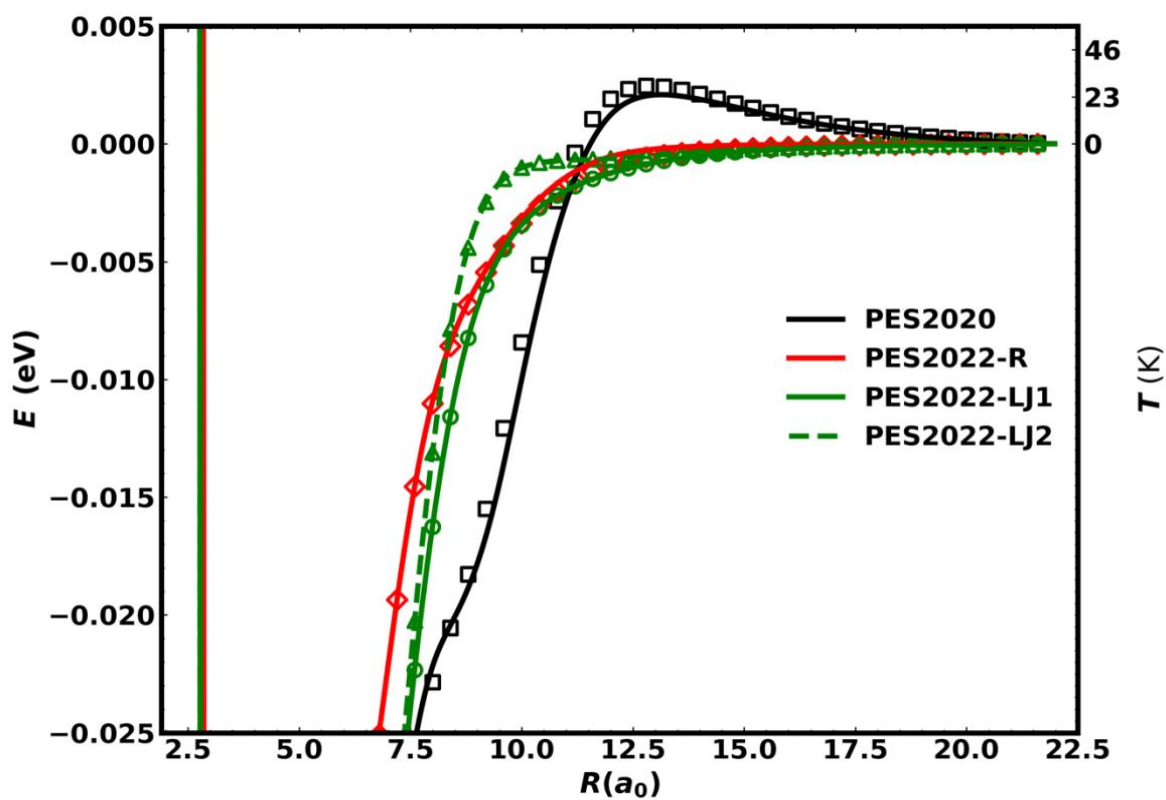

**Figure 6 :** PESs for the N + NO approach at a fixed value of r = 2.185 a0 and θ= 129.0° (symbols) and averaged over NO-separations between r = 2.10-2.29 $a_0$ in steps of 0.01 $a_0$ (solid lines). The distance R between the center of mass of N-O and N is scanned and the zero of energy is taken to match the N-O diatomic. PES2020 (black), PES2022-LJ1 (green), PES2022-LJ2 (green dashed), and PES2022-R (red). It is found that even with averaging, the spurious barrier on PES2020 persists whereas for all PES2022 there is no unphysical barrier at long range.

Various characteristics of the PESs (energies, equilibrium bond lengths and bond angles) are summarized in Table 2.

**Table 2** Minimum (MIN) and transition state (TS) for the $^3A''$ PESs considered in the present work calculated using the Nudged Elastic Band (NEB)[54, 55] method. Equilibrium NN and NO separations $r_e^{(NN)}$ and $r_e^{(NO)}$ in $a_0$, angle $\angle$(NON) in degrees, and energies $\Delta E_1$ (in eV) and $\Delta E_2$ (in kcal/mol) with respect to the N + N + O asymptote for the $^3A''$ PES. For PES2020, see Ref. 9.

|  | $r_e^{(NN)}$ | $r_e^{(NO)}$ | $\angle$(NON) | $\Delta E_1$ | $\Delta E_2$ |
|---|---|---|---|---|---|
| PES2020, PES2022-LJ1, PES2022-LJ2 | | | | | |
| MIN | 3.92 | 2.51 | 102.9 | -4.651 | -107.245 |
| TS | 3.97 | 2.36 | 102.0 | -4.629 | -106.746 |
| PES2022-R | | | | | |
| MIN | 3.92 | 2.51 | 102.9 | -4.639 | -106.981 |
| TS | 3.97 | 2.36 | 102.0 | -4.618 | -106.482 |

As none of the three PESs derived in this work present a barrier towards N approach, the calculated reaction rates continue to increase as the temperature falls below 75 K, unlike those derived from the PES2020 surface. At higher temperatures (~300 K), all PESs yield comparable rates consistent with experiment whereas towards lower temperature PES2022-R compares best with the measured rates from the present and earlier work. It should, however, also be noted that around 50 K the rate from PES2022-R flattens out and only continues to increase for PES2022-LJ1. This indicates that it is not only the spurious barrier at long range that influences the behaviour of $k(T)$ at low temperature, but also further details such as the slope of the PES at and beyond ~8 $a_0$.

    The two PES-correction schemes employed here have their own merits and disadvantages. Extrapolation based on explicit long-range interactions depends on the

availability and quality of the long-range coefficients, such as electrical moments or isotropic/anisotropic van der Waals coefficients.[39] As shown in Table 2, the geometries and minimum/transition state energies between the unmodified and modified PESs remain unchanged, which is advantageous. Completing and extrapolating an existing grid, on the other hand, does not depend on empirical expressions and availability of (experimentally measured) parameters but may be contingent on the location and density of missing points in the grid. Also, although the geometries of the minimum and transition state are the same as those of the original PES2020, the absolute energies change (by 0.25 kcal/mol) whereas the energy differences remain unaffected, see Table 2.

**5 Discussion and Conclusions**

The present experimental and theoretical results and the earlier measurements of Bergeat et al.[25] are in contrast to the majority of previous theoretical studies where the rate coefficients for the N + NO reaction were calculated to fall below 75 K (positive temperature dependence). Although the current experimental work does not extend to sufficiently low temperatures to unambiguously confirm the observed trend and the discrepancy with earlier calculations, the use of low and continuously decreasing rate coefficients for this reaction in current astrochemical databases (based on the trend displayed by these previous calculations) is debatable. A modified Arrhenius type fit to the present experimental data with the form $\alpha(T/300)^\beta \exp(-\gamma/T)$ (with $\gamma = 0$), yields the following expression, $k = (2.7 \pm 0.2) \times 10^{-11} (T/300)^{-(0.50 \pm 0.02)}$ cm$^3$ s$^{-1}$, which is valid over the range 50-296 K. Extrapolating this expression down to 10 K for interstellar modelling purposes yields a rate coefficient of $1.5 \times 10^{-10}$ cm$^3$ s$^{-1}$. This compares with the estimated value of the rate coefficient for this reaction at 10 K included in current astrochemical databases of $1.1 \times 10^{-11}$ cm$^3$ s$^{-1}$, an order of magnitude lower than the extrapolated value from the present work.[26]

Interestingly, the rate coefficients obtained here are more consistent with the estimated values used in earlier networks such as the Ohio State University astrochemical database (see Table 5 of Wakelam et al.[56]) where the rate coefficient for the N + NO reaction was also predicted to increase as the temperature falls reaching a value of $2.3 \times 10^{-10}$ cm$^3$ s$^{-1}$ at 10 K.

Another consequence of the use of an increased rate for the N + NO reaction at low temperature is the possible knock-on effect this might have on the rate coefficients for other reactions involving ground state atomic nitrogen. Indeed, the earlier low temperature kinetic study by Daranlot et al. [57] employed a relative rate technique to obtain rate coefficients for the N + OH reaction, based on the experimental study of the N + NO reaction by Bergeat et al. [25] as a reference. The same relative rate technique was then applied to determine rate coefficients for the N + CN, [6] N + CH, [58] N+ $C_2$ [59] and N + $C_2$N [60] reactions based on the rate coefficients for the N + OH reaction derived by Daranlot et al. [57] as a reference. In light of the present results, the rate coefficients for these reactions involving atomic nitrogen should be reviewed in future astrochemical modeling studies to determine whether their rate coefficients need to be reevaluated, supported and accompanied by simulations such as those carried out here and for other, related systems.[9, 49, 61-64]

In summary, this joint experimental and theoretical investigation of the N + NO reaction allowed us to determine rate coefficients for this process at room temperature and below (30 - 300 K). Experiments were conducted down to 50 K on a supersonic flow (Laval nozzle) type apparatus employing pulsed laser photolysis for the (indirect) production of ground state atomic nitrogen and pulsed laser induced fluorescence at VUV wavelengths to detect these atoms. In parallel, quasi-classical trajectory calculations were performed down to 30 K using modified PESs based on an earlier $^3$A'' global potential energy surface for $N_2$O in which a spurious barrier at long range was removed using two different procedures. The measured and calculated rate coefficients obtained here are in good agreement, displaying a monotonic *increase* as the temperature falls. While these results validate earlier kinetic experiments performed over a similar temperature range,[25] they disagree with the majority of previous theoretical studies that predict a *decrease* in reactivity as the temperature falls below 75 K.[8, 11, 12] The results of the QCT simulations based on all modified PESs, but in particular using PES2022-R, show favourable to excellent agreement with the present and previous experimental kinetics studies down to *T* = 50 K. Experiment was pivotal in reassessing the quality of PES2020 and led to critically investigate details on the ~50 K energy scale in the long range of the intermolecular interactions which guide the initial phase of a chemical reaction. The findings strongly suggest that the rates for the N + NO reaction could be considerably larger than those used in current astrochemical models. Consequently, the results and conclusions of models employing the currently recommended rate coefficients for this reaction should be reassessed in light of the present results.

The present work demonstrates that the interplay of experiment and simulations provides a stringent test of the intermolecular interactions and leads to improved, global and reactive PESs for future simulation studies. In addition, it is demonstrated that for reliable *T*-dependent information, such as reaction rates or cross sections, at the temperatures relevant to astrophysical modeling (T ~ 10 K), dedicated experimental and computational studies are required.

**Acknowledgements**

K. M. H. acknowledges support from the French program ''Physique et Chimie du Milieu Interstellaire'' (PCMI) of the CNRS/INSU with the INC/INP co-funded by the CEA and CNES as well as funding from the ''Programme National de Planétologie'' (PNP) of the CNRS/INSU. The research in Basel was supported by the AFOSR and the University of Basel, which both are gratefully acknowledged.

**Footnotes**

Electronic supplementary information (ESI) available: van der Waals N + NO long-range potential, RKHS evaluation at different angular cuts for short (*R*< 7.0 $a_0$) and long (*R*>= 7.0 $a_0$) ranges, contour plots of the RKHS evaluated energy surface at fixed equilibrium N-O separation of *r*=2.185 $a_0$ and experimental and calculated rate coefficients for the N + NO reaction.

Electronic Supplementary Information for

**Low-Temperature Kinetics for the N + NO reaction: Experiment Guides the Way**

Kevin M. Hickson,*,[1] Juan Carlos san Vicente Veliz,[2] Debasish Koner[2,3] and Markus Meuwly*,[2,4]

[1]Univ. Bordeaux, CNRS, Bordeaux INP, ISM, UMR 5255, F-33400 Talence, France
[2]Department of Chemistry, University of Basel, Klingelbergstrasse 80, CH-4056 Basel, Switzerland.
[3]Department of Chemistry, Indian Institute of Technology Hyderabad, Sangareddy, Telangana 502285, India.
[4]Department of Chemistry, Brown University, Providence, RI 02912, USA

**Figures S1-S3 and Table S1**

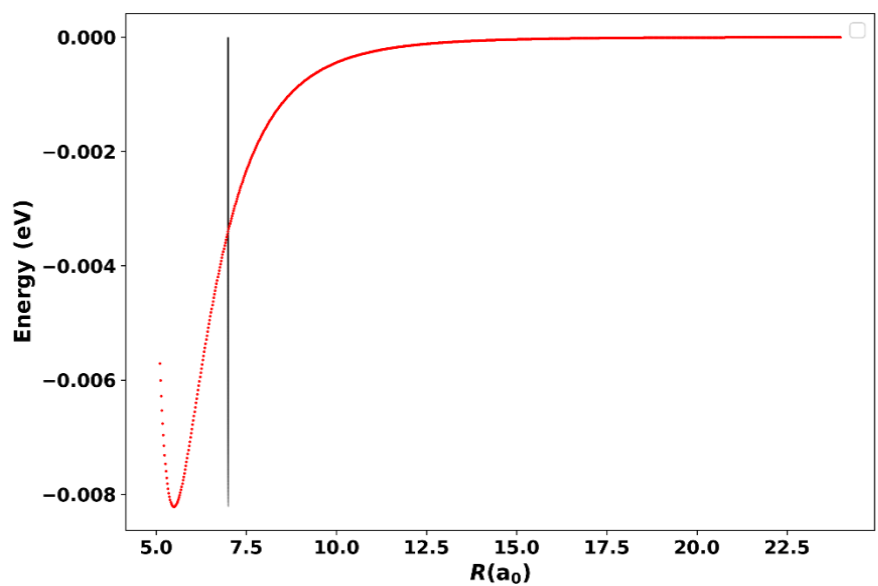

**Figure S1** van der Waals NO + N long-range potential $V(R)$. Ratio between $V(R = 9)/V(R = 7)$ and $V(R = 12)/V(R = 7)$ are 0.2168 and 0.0380 respectively.

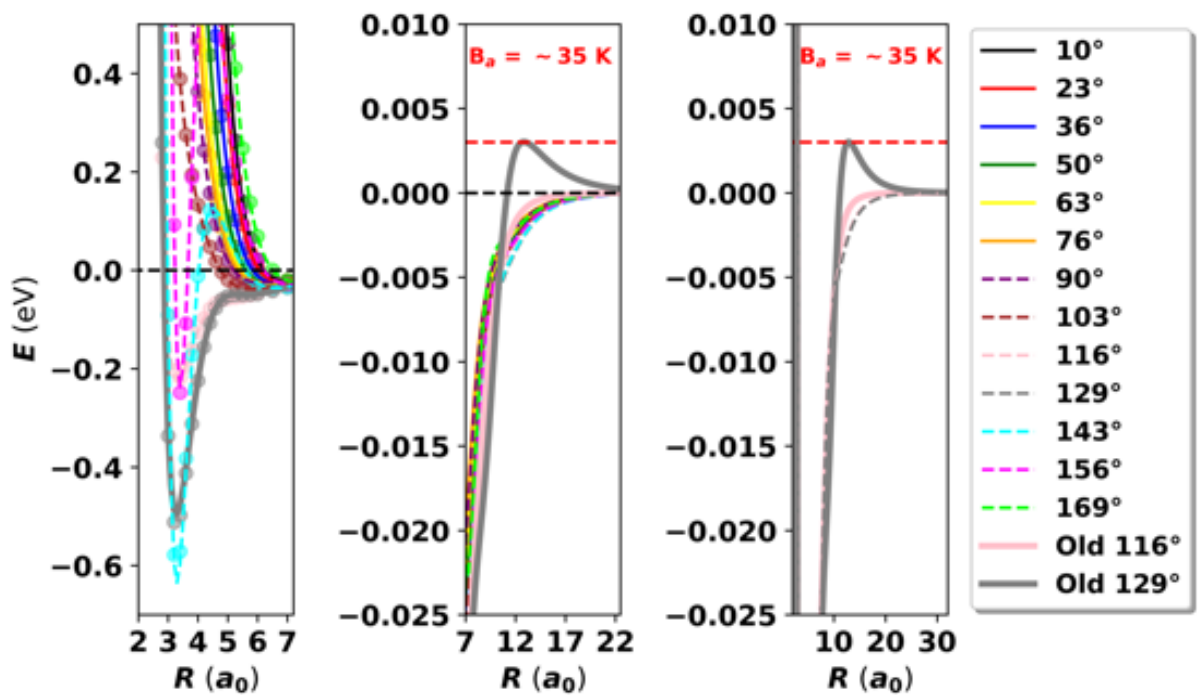

**Figure S2** RKHS evaluation at different angular cuts for short ($R< 7.0$ $a_0$) and long ($R\geq 7.0$ $a_0$) ranges. **Top panel:** The evaluations are at fixed $r=q^-=2.100$ $a_0$ (inner turning point), **Middle panel:** $r=r_{eq}=2.185$ $a_0$, **Bottom Panel:** $r=q^+=2.280$ $a_0$ (outer turning point). Points represent ab-initio reference (symbols), RKHS evaluation with PES2022-LJ1 (solid/dash lines), and RKHS evaluation with PES2020 (thick solid line) (116° and 129°).

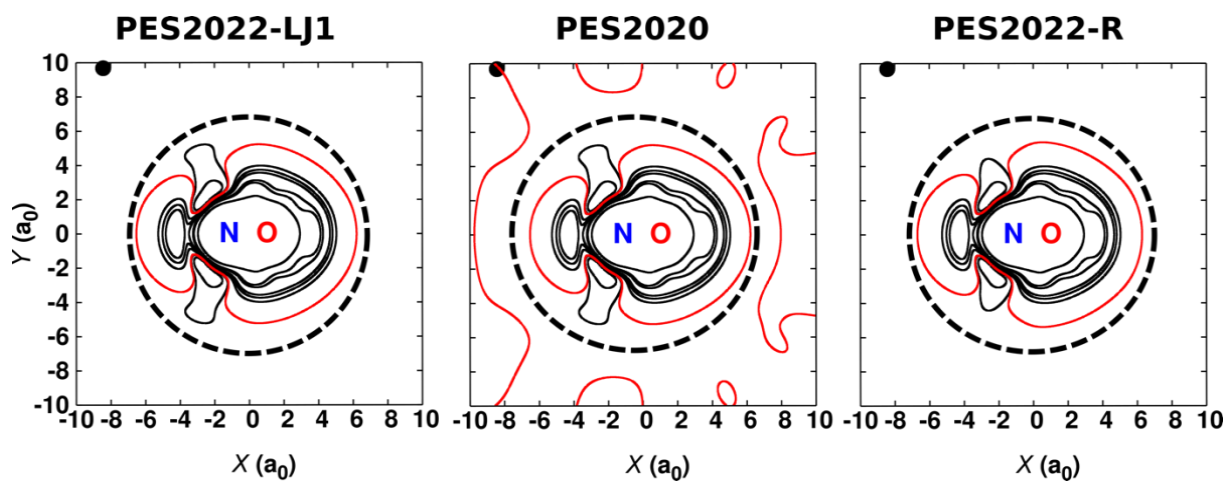

**Figure S3** Contour plots of the RKHS evaluated energy surface at fixed equilibrium N-O separation of $r=2.185$ $a_0$. **Left panel:** for PES2022-LJ1, **Middle Panel:** PES2020, and **Right Panel**: PES2022-R. Contour lines are shown for energies [-0.3, -0.05, 0.45, 0.71, 0.96, 2.0, 2.5, 6.7] eV (black solid lines) and energy is reported with respect to dissociation to N + NO ($r_{eq}$).

To highlight the pronounced (unphysical) anisotropy present in PES2020 a selected isocontour at 0.0031 eV= 36 K (red solid line) is drawn to indicate the barrier at $\theta$=129.0º, $R$=12.85 $a_0$ in the XY representation (-8.43,9.69) $a_0$. The solid black circle indicates a reference geometry to highlight that PES2020 contains an isocontour with positive energy at large separation $R$ whereas PES2022-LJ1 and PES2022-R do not. The black dashed line is a circle with a radius of 7.0 $a_0$ which contains all the data we have retained from the MRCI+Q calculations.

**Table S1** Experimental and calculated rate coefficients ($\times 10^{-11}$ cm$^3$ s$^{-1}$) for the N($^4$S) + NO(X$^2\Pi$) → O($^3$P) + N$_2$($X^1\Sigma_g^+$) reaction.

| $T$ (K) | Expt. | PES2020 (HB) | PES2022-R (HB) | PES2022-R (GB) | PES2022-LJ1 (HB) | PES2022-LJ2 (HB) |
|---|---|---|---|---|---|---|
| 300 |  | 3.33 | 2.95 | 3.13 | 3.73 | 3.60 |
| 296 | 2.81 | 3.41 | 2.98 | 2.92 | 3.75 | 3.62 |
| 200 |  | 3.92 | 3.71 | 3.82 | 4.78 | 4.44 |
| 177 | 3.50 | 4.19 | 3.93 | 4.21 | 5.17 | 4.75 |
| 127 | 4.00 | 4.59 | 4.69 | 4.41 | 6.42 | 5.69 |
| 100 |  | 4.88 | 5.37 | 5.41 | 7.42 | 6.35 |
| 75 | 5.27 | 5.04 | 6.09 | 6.24 | 8.62 | 7.07 |
| 50 | 6.64 | 4.83 | 6.80 | 6.54 | 9.97 | 7.64 |
| 30 |  | 4.14 | 6.92 | 6.80 | 10.88 | 7.72 |